\documentclass{article}

\usepackage{arxiv}

\usepackage[utf8]{inputenc} 
\usepackage[T1]{fontenc}    
\usepackage{hyperref}       
\usepackage{url}            
\usepackage{booktabs}       
\usepackage{amsfonts}       
\usepackage{nicefrac}       
\usepackage{microtype}      
\usepackage{lipsum}		
\usepackage{graphicx}
\usepackage{natbib}
\usepackage{doi}
\usepackage[section]{algorithm}
\usepackage[noend]{algpseudocode}
\usepackage{subfig}

\title{Ontological model identification based on data from heterogeneous sources}

\author{ \href{https://orcid.org/0000-0000-0000-0000}{\includegraphics[scale=0.06]{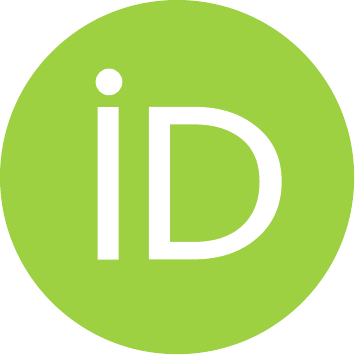}\hspace{1mm}Egor Shikov}\thanks{Use footnote for providing further
		information about author (webpage, alternative
		address)---\emph{not} for acknowledging funding agencies.} \\
	ITMO University\\
	St Petersburg, 49 Kronverksky pr. \\
	\texttt{egorshikov@itmo.ru} \\
	\And
	\href{https://orcid.org/0000-0000-0000-0000}{\includegraphics[scale=0.06]{orcid.pdf}\hspace{1mm}Danila Vaganov} \\
	ITMO University\\
	St Petersburg, 49 Kronverksky pr. \\
	\texttt{vaganov@itmo.ru} \\
	\And
	\href{https://orcid.org/0000-0002-9369-7104}{\includegraphics[scale=0.06]{orcid.pdf}\hspace{1mm}Anton Lysenko} \\
	ITMO University\\
	St Petersburg, 49 Kronverksky pr. \\
	\texttt{blinkop@gmail.com} \\
	\And
	\href{https://orcid.org/0000-0000-0000-0000}{\includegraphics[scale=0.06]{orcid.pdf}\hspace{1mm}Alexander Kalinin} \\
	ITMO University\\
	St Petersburg, 49 Kronverksky pr. \\
	\texttt{amkalinin@niuitmo.ru} \\
}

\renewcommand{\shorttitle}{\textit{arXiv} Template}

\hypersetup{
pdftitle={Ontological model identification based on data from heterogeneous sources},
pdfsubject={q-bio.NC, q-bio.QM},
pdfauthor={A .Kalinin et al.},
pdfkeywords={ontological models, knowledge graphs, relationship extraction, heterogenous data integration},
}

\begin{document}
\maketitle

\begin{abstract}
The development of a company often entails the emergence of autonomous data sources with different structural and technological organization. This can lead to the inability of data analysis at a high level and a violation of the integrity and reliability of data within the organization, hindering the adoption of high-quality decisions and further development of the company. This problem can be solved by implementing a higher abstraction, representing heterogeneous organization data in a single space by combining them into a single knowledge graph. We propose a framework capable of autonomous construction of an organization's knowledge graph based on semi-structured data from various sources by finding links between sources based on data with an arbitrary structure, and combining document collections into single entities. The results of tests show the applicability of the developed approach for constructing a knowledge graph based on partially-structured data from various sources and the high efficiency of the approach based on the metrics of completeness of data storage subsystems coverage (11 out of 11) and filtering false connections (there are only 2.5 connections of collections with neighbors on average in the final graph).
\end{abstract}

\keywords{ontological models, knowledge graphs, relationship extraction, heterogenous data integration}

\section{Introduction}
With the development of the organization, the opening of new departments, involving various counterparties, as well as the use and development of new information systems, designed to improve autonomy and efficiency of individual departments, the diversity and number of independent data sources increasing. Organizations are not always able to implement and maintain successful data management concept (Data Governance) in a timely manner, leading to low quality of evidence and the inability to assess the real situation within the company operations and make high-level decisions. The lack of completeness, consistency and reliability of information at the level of the entire organization can lead to loss of profit and the impossibility of further development of the company. 

Thus, at the level of each individual department, independent data warehouses are developed with an arbitrary structure and technologies used, making the process of combining all the organization's data into a single system expensive and involves global changes in business processes and technical equipment of existing information systems. 

The problem of data integration of heterogeneous structures from various sources differing in technological equipment and semantics can be solved by the introduction of higher-level abstractions representing heterogeneous data organization in a single space by combining them into a single knowledge graph. In such graph the nodes are entities, which are grouped based on multiple tables collections (including from different sources) that describe an individual object within the organization, and edge is the fact of a relationship between such objects, denoted by the semantics, extracted from the data: inclusion, intersection, inheritance, etc.

In this project existing solutions for auxiliary graph building knowledge of the organization on the basis of data from different sources (p. 1) are researched, and an approach to solving this problem is developed. The basis of this approach is a set of methods and algorithms, including filtering and pre-processing semi-structured data (section 3), the algorithm of searching links between sources based on the data with any structure (**4), and an algorithm of combining the collections of documents in a single entity (**5). This approach is implemented as a software library in Python 3.9 and tested for cross-domain data organization provided by the customer. The results of testing show the applicability of the developed approach for constructing a knowledge graph based on partially structured data from various sources and the high efficiency of the approach based on the metrics of completeness of data storage subsystems coverage (11 out of 11) and filtering false connections (there are only 2.5 connections of collections with neighbors on average, in the final graph).

\begin{figure}[h]
    \centering
    \includegraphics[width=\linewidth]{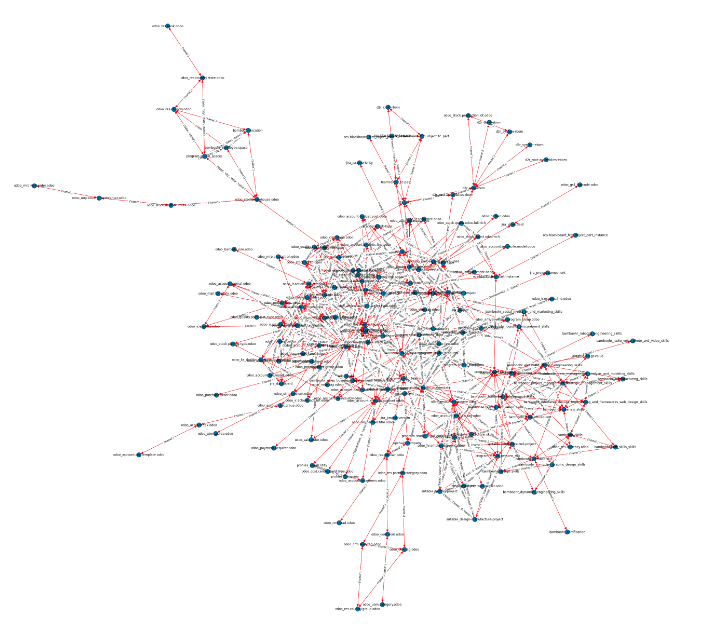}
    \caption{Example graph.} \label{fig:graph}
\end{figure}

\section{Theoretical background and related work}
\subsection{Knowledge graphs and ontologies}
Despite the intensive development of the region, there is no commonly accepted definition of the term "graph knowledge" \citep{Hogan2021}. Most definitions note that this is a graph representation of real-world objects and their relationships. In general, the knowledge graph includes two components: data and a schema.

Data in the knowledge graph is represented as triplets: < subject, predicate, object>. A subject can only be an entity and acts on an object through some predicate (relation). An object in a triplet can be either an entity or an attribute (literal). In the first case, the predicate is called an object predicate, and in the second case, it is called a literal predicate.

To add semantics to the accumulated data, the knowledge graph is supplemented with an ontology. An ontology represents the schema of data and its semantics. An ontology consists of classes (entity types) and relationships (edge types). The ontology also contains axioms about classes (TBox, terminology box) and relations (RBox, relation box). Axioms about classes allow you to define a hierarchy of classes (equivalence and subordination relations). Axioms about relations allow you to define the hierarchy of relations, as well as properties such as transitivity, reflexivity, symmetry, incompatibility, and so on.

Object classes and literal types define the domain space of predicate definitions and values. It is the set of classes to which a subject can belong in a triplet with a given predicate. Value domain is a set of classes or literal types that a triplet object can belong to.

ABox (assertion box) contains instances (entities) classes and relationships, i.e. nodes and edges. An ABox can also include a statement about the equivalence of entities.

Most ontology description languages are based on mathematical logics. The more expressive the logic, the higher the algorithmic complexity. Among the widely used languages for describing ontologies, we can distinguish RDF, RDFS, OWL 1, OWL 2. The latter also has several profiles that have different levels of expressiveness.

The features of TBox and RBox determine the expressiveness of an ontological language. Moreover, the expressiveness of the language depends on the ability to create composite (complex) classes. Composite classes can be described by expressions consisting of simple classes and relationships.

You can follow two paradigms while working with knowledge graphs: the "open world hypothesis" and the "closed world hypothesis". The first, Open World Assumption (OWA), assumes that the absence of information in the database does not mean that it is true or false. The second hypothesis, the Closed World Assumption (CWA), in contrast, sets a strict condition that all true information is stored in the database.

\subsection{Integrating structured data into knowledge graphs}

Two NoSQLdatabases are considered in \citep{Cure2013}: document-oriented (MongoDB) and columnar (Cassandra). A "local" ontology is extracted from each database. Then, a common "global" ontology is constructed by comparing the classes of the two ontologies. The authors use a virtual approach to make queries to the resulting ontology. This approach implies that queries written in one language are translated into the languages of the used DBMS. The general scheme of the solution is shown in Figure 1.1.
 

To extract an ontology from data, the authors developed rules for converting JSON (MongoDB) documents to RDF.
\begin{itemize}
    \item Each collection in the database is treated as a class.
    \item Each field in a document is a relation whose domain space is the document class.
    \item Fields with standard types (strings, numbers, boolean values) are converted to literal relations with the value range corresponding to the value type.
    \item If the field values are an array, then there are two possible options.
    \begin{itemize}
        \item If IDs of other entities are listed in the array, then an object relation is added. The value domain of this relation is the classes of entities, whose IDs are listed in the array.
        \item If an array contains enumeration values that have a finite set of possible values, then each such element of this array is considered as a class. An object relation is added whose value domain is the classes specified in the array. 
    \end{itemize}
\end{itemize}

 Rules for converting JSON documents into an ontology are also proposed in papers \citep{Abbes2015, Abbes2017}.The data source is the NoSQLdatabase MongoDB :
 \begin{itemize}
     \item 	Each collection in the database is treated as a class.
     \item Each field in a document is a relation whose domain space is the document class.
     \item Fields with standard types (strings, numbers, boolean values) are converted to literal relations with the value range corresponding to the value type.
     \item If the field value is a nested object, then an object relation is created. The domain space of the relationship is the class of the current document. A new class is extracted from the nested object, and the field name is assigned to it. This class is defined as the value area of the selected relationship. The relationship name is formed from the field name and the "has" prefix.
     \item Similarly to item "d", transformation is also applied for "foreign keys" that are supported in MongoDB using the "DB Refs" mechanism. In this case, the value domain of the selected relationship is the document class that the field points to.
     \item If there are fields named "child" or "parent" that point to other documents, then subordination relationships are created, i.e. a class hierarchy.
     \item All field values are converted to entities.
     \item An axiom of equivalence (owl:equivalentClass) is added between two classes if the entities of these classes completely coincide. Otherwise, the incompatibility axiom is added (owl: disjointWith).
     \item For each object relation, an inverse relation is added, the value domain and domain space of which are equal to the value domain and domain space of the original relation, respectively. The relationship name is formed from the prefix "belongs\_to" and the name of the original relationship without "has".
     \item For fields whose values are represented by an array, there are restrictions on the number of possible edges: owl: maxCardinality, owl:minCardinality, owl:cardinality.
 \end{itemize}
 
An arbitrary file in JSON format is considered as the data source in \citep{Sbai2019}. At the first stage, the class hierarchy is calculated. To do this, the authors analyze the field names of nested objects. The parent class is selected from two objects if some of their fields intersect. At the second stage, the authors apply the transformation rules:
\begin{itemize}
    \item Each object in the JSON file is converted to a simple class.
    \item Fields with standard types (strings, numbers, boolean values) are converted to literal relations with the value range corresponding to the value type. The relationship name is formed from the "has" prefix, the field name, and the class name from the definition area.
    \item If the field value is a nested object, then an object relation is created. The domain space of the relationship definition is the class of the current document. A new class is extracted from the nested object, and the field name is assigned to it. This class is defined as the value area of the selected relationship. The relationship name is formed from the field name and the "has" prefix. The root object is named "Class1". Other objects that cannot be named are also referred to as "ClassN", where N is the class number.
\end{itemize}
	
\subsection{Integrating unstructured data into knowledge graphs}

Despite the fact that there is a lot of data that is stored in a format that has some structure, most of the information is presented in the form of arbitrary text. In this regard, it makes sense to extract facts directly from symbolic representations. Converting text into a knowledge graph can be divided into four stages.
\subsubsection{Preprocessing}
Original text goes through standard procedures used in natural language processing:
\begin{itemize}
    \item tokenization (highlighting individual words);
    \item automatic morphological markup (part-of-speech detection);
    \item building a dependency tree where leaf nodes contain individual words that form phrases or whole sentences;
    \item determining the meaning of a word.
\end{itemize}

\subsubsection{Named Entity Recognition (NER)}
Named entity recognition consists of identifying references to entities of a certain class (people, organizations, locations, etc.) in the text. There are several types of RICE. A number of solutions use lexical features (morphology, dependency tree) and reference books (lists of commonly-used names, countries, and businesses). Learning methods with a teacher \citep{Lample2016} require preliminary data markup. Since the creation of the markup is a costly process, methods used procedure of bootstrapping \citep{Gupta2014, Nakashole2013, Yogatama2015}, which increases initially a small training sample. Learning methods with minimal teacher involvement \citep{Ling2012, Ren2015} using well-known entities of the knowledge graph as the initial training sample in order to identify similar entities in the text. Methods based on manual rules are also used \citep{Chiticariu2018, Kluegl2009}. Such methods are better controlled and their behavior is more predictable \citep{Chiticariu2013}.

The selected entities from the text must be linked to those that already exist in the knowledge graph. The task can be considered as a ranking, where each mention from the text is compared with all entities \citep{Moro2014, Wu2018}. The task is complicated by the fact that the reference may occur in different contexts, and also by the fact that one entity may be referred to differently in the text.

\subsubsection{Extracting relationships}
After recognizing entities, the search is performed for relationships that can be associated with these entities. A "closed configuration" implies that the set of relationships you are looking for is defined in advance. Methods of this class can be based on rules, use supervised learning \citep{Roller2018}, bootstrapping \citep{Bunescu2005,  Etzioni2004}, training with minimal involvement of teachers \citep{Smirnova2018}. In an "open configuration", the set of relationships is not defined in advance, but is extracted entirely from the text. In this case, unsupervised learning is used.

The described division into stages is not strict. Some methods can perform part of the steps together in order to improve the quality of the result.

\subsection{Ontology matching methods}

Creating the most accurate and comprehensive knowledge base requires combining several different sources. However, the data in the sources may overlap, so the resulting graph will also contain duplicate information. Comparison of different graph knowledge consists in definition of the entities, which is one of the real-world object \citep{Nentwig2017, Sun2020,  Zhao2020}.

The problem of identifying duplicates was initially solved for records in relational database tables. To map vertices, feature description made up of table fields is used. Having a set of matched pairs of vertices and their feature descriptions, you can train standard machine learning algorithms for the classification problem in the supervised learning. This approach can also be generalized to graphs by including attributes from adjacent vertices in the table.

Since checking all possible pairs of entities is a resource-intensive operation, a set of candidate entities from the target source is formed for each entity from the source source. This procedure is called blocking. The essence is to determine the most likely possible duplicates, the number of which is much smaller than the size of the entire set of entities. The similarity of records can be determined in various ways. The simplest and most widely used method is calculating the proximity of entity names. Framework Magellan \citep{Konda2016} provides a set of tools for composing blocks and matching records using various machine learning algorithms (for example, decision trees, random forest, gradient boosting). DeepMatcher \citep{Mudgal2018} allows to obtain higher quality mapping through the use of deep learning models.

\begin{figure}[h]
	\centering
	\subfloat[\label{sub:hours}]{\includegraphics[width=0.45\textwidth]{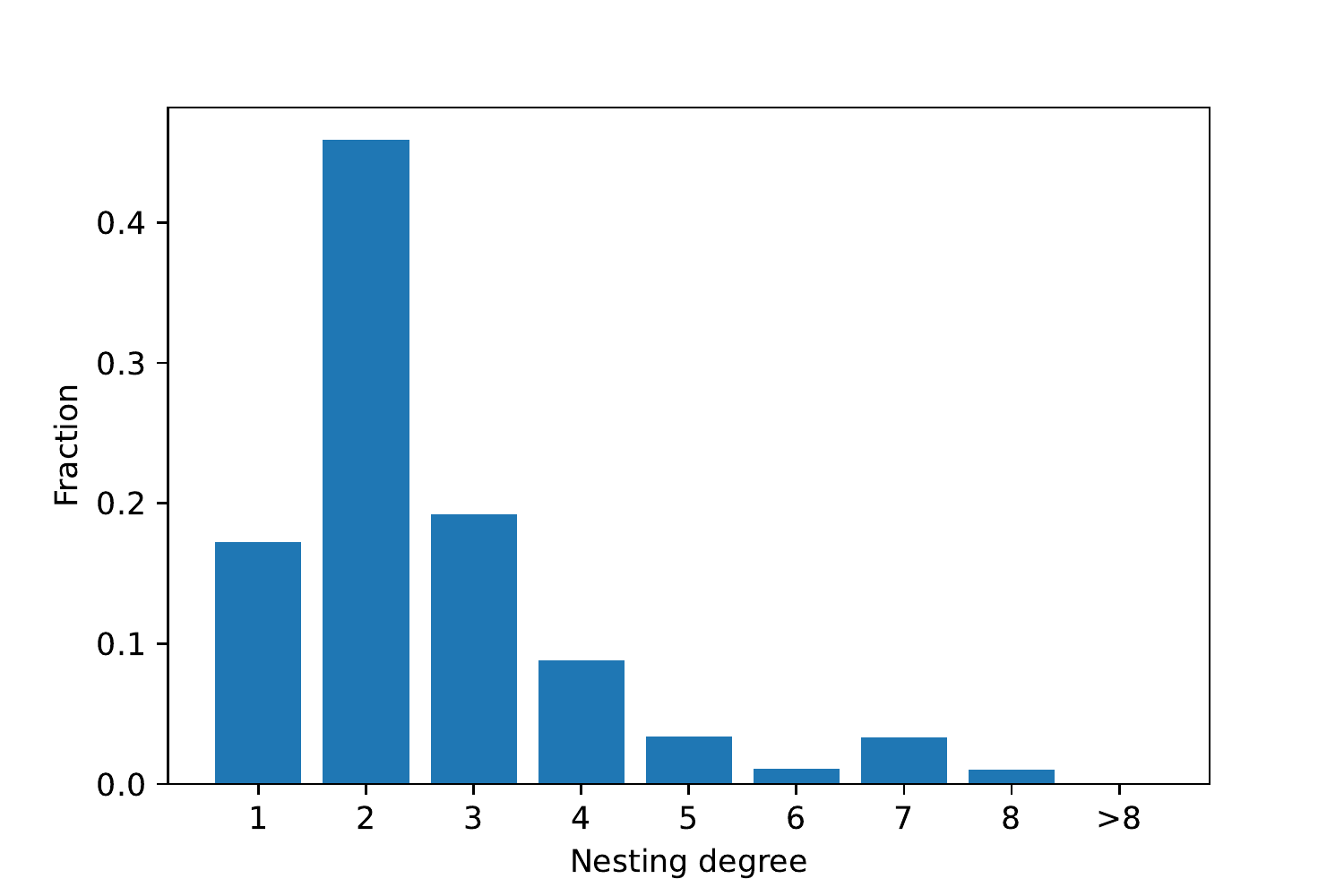}}\qquad
	\subfloat[\label{sub:weekdays}]{\includegraphics[width=0.45\textwidth]{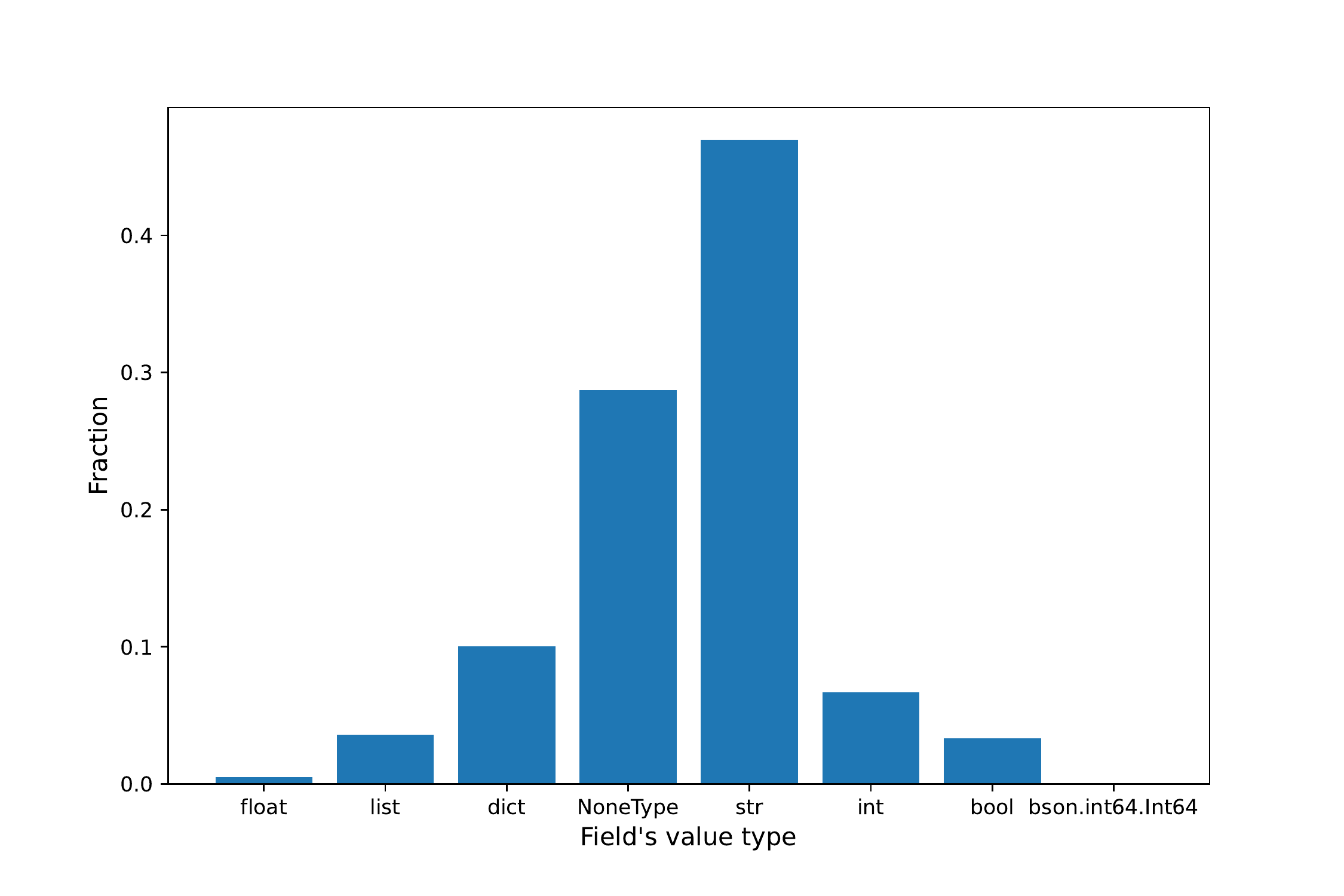}}\qquad
	\caption{Distribution of the nesting degree of objects (left); distribution of object types in all collections (right)}
	\label{fig:ncoll}
\end{figure}


\section{Methodology}
\subsection{Preprocessing and filtering}
Relational databases maintain relationships between tables using primary and foreign keys. The first one is the ID of a specific record in the table, and the second one indicates the ID of a record from another table. However, NoSQL databases either have no such mechanisms, or these mechanisms are rarely used by system developers due to the lack of a strict data schema. Moreover, it is often more convenient and efficient to store part of an external document as a nested object in the current document. This raises the problem of recognizing keys among document fields. The names of documents and their temporal component-fields that store dates are also important information in addition to keys.

\begin{algorithm}[h]
\caption{Keys, names and dates searching }\label{algo1}
\begin{algorithmic}[1]\small
    \Procedure{FindKeys}{document,find\_primary}
    
        \State primary\_list $\gets \emptyset$
        \State foreign\_list $\gets \emptyset$
        \State name\_list $\gets \emptyset$
        \State date\_list $\gets \emptyset$
        
        \For{field \textbf{in} document}
        \If{find\_primary \textbf{and} (($w \in W$ \textbf{in} field.name) \textbf{or} (field.value \textbf{is} hash))}
            \State primary\_list $\gets$ primary\_list $\cup$ \{ field\}
        \Else
        \State f, n, d $\gets$ \textsc{ExploreEmbedded}(field)
        
        \State foreign\_list $\gets$ foreign\_list $\cup$ f
        \State  name\_list $\gets$ name\_list $\cup$ n
        \State  date\_list $\gets$ date\_list $\cup$ d
        \EndIf
        \EndFor
        
        \If{find\_primary \textbf{and} len(primary\_list) $> 0$}
        \State primary\_key $\gets$ choose field from primary\_list, such that field.name  $w \in W$ (according to the priority)
        \State foreign\_list $\gets$ foreign\_list $\cup$ primary\_list \textbackslash \{primary\_key \} 
        
        \EndIf
        \If{find\_primary \textbf{and} len(primary\_list) $ = 0$} 
        \State primary\_key $\gets$ try to find field in foreign\_list, such that field.name includes $w \in W$ (according to the priority)
        
        \State foreign\_list $\gets$ foreign\_list \textbackslash \{primary\_key \} 
        \EndIf
        \If{find\_primary \textbf{and} \textsc{KeyIsComposite}(primary\_key)} 
        \State field.IdType = COMPOSITE
        \EndIf
        \If{find\_primary \textbf{and} \textsc{KeyIsMultiRef(primary\_key)}}
        \State field.IDType = MANY
        \EndIf
        \State \Return primary\_key, foreign\_list, name\_list, date\_list
    \EndProcedure

 \end{algorithmic}
\end{algorithm} 

\begin{algorithm}[h]
\caption{Keys, names and dates searching in embedded documents}\label{algo2}
\begin{algorithmic}[1]\small
    \Procedure{ProcessEmbedded}{field}
    
        \State foreign\_list $\gets \emptyset$
        \State name\_list $\gets \emptyset$
        \State date\_list $\gets \emptyset$
        
        \If{(($w \in W$ \textbf{in} field.name) \textbf{or} (field.value \textbf{is} hash))}
            \State foreign\_list $\gets$ foreign\_list $\cup$ \{ field\}
        \ElsIf{field.name = 'name'}
        \State name\_list $\gets$ name\_list $\cup$ \{field\}
        \ElsIf{field.value can be converted to the date}
        \State date\_list $\gets$ date\_list $\cup$ \{field\}
        \ElsIf{field.value \textbf{is} an array}
        \If{\textsc{KeyIsComposite}(field.value)}
        \State field.IdType = COMPOSITE
        \State \Return \{field.value\}, name\_list,date\_list
        \EndIf
        \If{\textsc{KeyIsMultiRef}(field.value)}
        \State field.IdType = MANY
        \State \Return \{field.value\}, name\_list,date\_list
        \EndIf
        \For{element \textbf{in} field.value}
        \State f, n, d $\gets$ \textsc{ProcessEmbedded}(element)
        \State foreign\_list $\gets$ foreign\_list $\cup$ f
        \State name\_list $\gets$ name\_list $\cup$ n
        \State date\_list $\gets$ date\_list $\cup$ d
        \EndFor
        \ElsIf{field \textbf{is} embedded document}
        \State p, f, n, d $\gets$ \textsc{FindKeys}(field, False)
        \State foreign\_list $\gets$ foreign\_list $\cup$ f
        \State name\_list $\gets$ name\_list $\cup$ n
        \State date\_list $\gets$ date\_list $\cup$ d
        \EndIf
        
        \State \Return foreign\_list, name\_list,date\_list
    \EndProcedure

 \end{algorithmic}
\end{algorithm}

\begin{algorithm}[h]
\caption{Composite key determination}\label{algo3}
\begin{algorithmic}[1]\small
    \Procedure{KeyIsComposite}{array}
    \If{len(array)$= 2$ 
    
    \State \textbf{and} (elements do not contain documents or arrays)
    
    \State \textbf{and} (array[0] is integer)
    \State \textbf{and} (array contains at least one string value)}
    \State \Return True
    \EndIf
    \State \Return False
    \EndProcedure
 \end{algorithmic}
\end{algorithm} 

\begin{algorithm}[ht]
\caption{Multiple key determination}\label{algo4}
\begin{algorithmic}[1]\small
    \Procedure{KeyIsMultiRef}{array}
    \If{len(array)$> 0$ \textbf{and} (all elements are hash-sum)}
    \State \Return True
    \EndIf
    \State \Return False
    \EndProcedure
 \end{algorithmic}
\end{algorithm}

In the MongoDB database, documents are divided into collections. We define the schema of a collection based on its first non-empty document. Each field in the document has a name and value. To identify the key by the field name, an ordered set of keywords  (for example, "Id", "uid" ,etc.)., whose order means that the word takes precedence over others is introduced. The proposed method recursively traverses the document using algorithm \ref{algo1}, which accepts the document as input and checks all fields for belonging to the primary / foreign key, name or date. An auxiliary algorithm has been developed for processing nested objects and arrays \ref{algo2}. To solve the problem, it is necessary to determine whether the key is composite or multiple (it refers to several documents at once). An algorithms \ref{algo3} and \ref{algo4} have been developed for this purpose, respectively.

The developed method is tested on all collections of the provided data. As you can see in Figure XX, the primary key was extracted for most collections. Moreover, the portion of "covered" collections for each source (collection set) is close to one. However, most of the extracted keys are integers (Figure XX).

Figure XX shows the distribution of extracted foreign keys by type. In this case, composite keys, represented by  two-element arrays (e.g., [815, “John Smith”]), and multiple keys, e.g. on several papers (e.g. [“5f7a09de6f6f4d1c8ca390e0”, “5f7a09dd7b6f4d1c8ca290f2”, “5f7a05dd1b6f4d1c8ca390ec”]) are more common. As long as different collections can have the same fields, the distribution of unique keys is analysed (figure XX). In this aspect, the fraction of numeric, string, and boolean values increases and becomes closer to the fraction of composite keys and arrays.

\subsection{Finding links between sources}

\subsubsection{Search}

The problem of finding links between sources is solved by searching for common identifiers among pairwise intersections of all collections. However, the difficulty lies in the fact that a large number of identifiers are not unique.

A typical example to illustrate this situation is data uploaded from a PostgreSQL-based database without additional metadata. In this case, identifiers are integer keys with a starting point at 0. In this case, a large number of false matches occur. A similar scenario is possible for other sources if the identifiers are names or other short string literals.

To solve this problem, it was proposed to construct a metric based on data on key intersections. Several options were suggested:

\begin{equation}
    N(U, V) = |U \cap V|
\end{equation}
\begin{equation}
    J(U, V) = \frac{|U \cap V|}{|U \cup V|}
\end{equation}
\begin{equation}
    I(U, V) = \frac{|U \cap V|}{min(|U|,|V|)}
\end{equation}
\begin{equation}
    PMI(U, V) = log(\frac{p(U, V)}{p(U)p(V)})
\end{equation}

\begin{figure}
    \centering
    \includegraphics[width=0.6\linewidth]{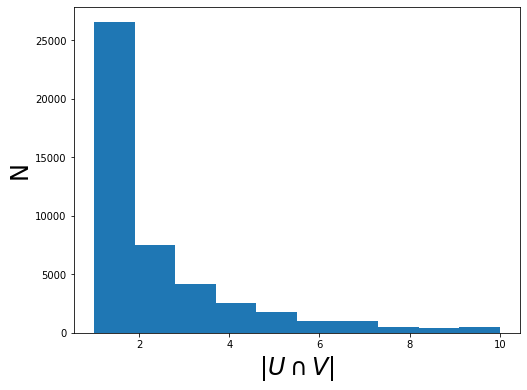}
    \caption{Distribution of the number of elements in the intersection of collections.}
    \label{fig:4_1}
\end{figure}

The first option is simply the number of intersections between the sets of keys contained in the $U$ and $V$ collections. However, this metric is difficult to use for filtering, as the number of intersections can vary greatly between pairs of collections. An exception is the use of the criterion $N(U,V)>1$, since the use of relationships does not make sense when traversing only one key. At the same time, as you can see in Figure \ref{fig:4_1}, such intersections significantly exceed all others, so it is advisable to use such a simple criterion.

The second option is a Jaccard coefficient. The third option is designed to solve the problem of a low value of Jaccard coefficient in the case of a connection between a collection with high and low cardinality.

The fourth option is a modified point mutual information between the sets $U$ and $V$. In this case, the probabilities of choosing a given pair of collections with the key $z$ are used, expressed in counting the number of collections with the key $z$. This approach reduces the impact of very frequent keys that occur in a large number of collections, such as "0".

\subsubsection{Filtering}

To compare the presented metrics, the following criterion was used: multiple intersections were filtered for each of the metrics and the dependence of the number of vertices on the number of edges for each filtration was constructed. Then the following assumptions were used:

\begin{itemize}
  \item All collections must be connected to at least one neighboring collection. Reducing the number of vertices during filtering means that real links are cut off incorrectly.
  \item For an ideal metric, the condition $M(x,y) < M(i, j)$ holds for all $x, y, i, j$, where the pair $(x,y)$ represents a real relationship, and the pairs $(i,j)$ are false – positive.
\end{itemize}

In this case, it is reasonable to expect that the "ideal" filtering criterion allows you to reduce the number of edges without reducing the number of vertices.

\begin{figure}
    \centering
    \includegraphics[width=0.6\linewidth]{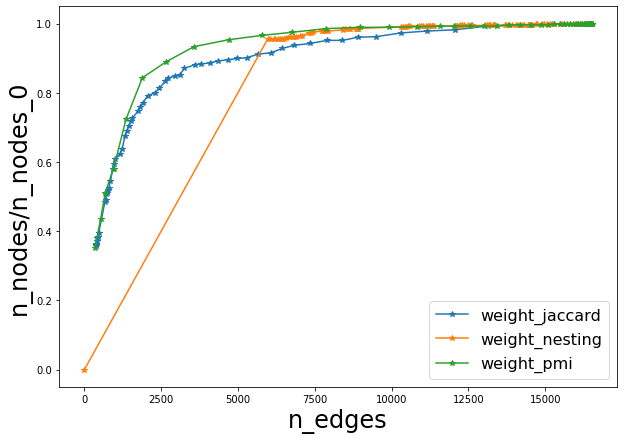}
    \caption{Dependence of the share of collections, involved in the graph, on the number of edges in this graph (for integer keys and for various metrics.}
    \label{fig:4_2}
\end{figure}

The calculation results are shown in Figure \ref{fig:4_2}. You can see that the metric based on mutual entropy shows the best result, since it has the largest area under the curve. This means that the specified metric allows us to remove the maximum number of false-positive links without significantly reducing the true-positive links.

\subsection{Entities integration}

Different collections may contain information from different sources/tables of any DBMS, but at the same time two different collections can describe properties or events of the same entity. Also, two different collections can represent the same information if these collections are built on different versions of the same database. Here we describe the developed method of combining such collections in entities in order to make the graph more compact and more informative.

Using the algorithm for finding links between sources and identified primary keys in these sources, we can construct a matrix $A \in \mathbb{R}^{n \times n}$, where $n$ is the number of collections, and the $A_{ij}$ occurrence of $A$ is the maximum metric value of the intersection of a pair of primary keys between $i$-th and $j$-th sources. The degree of coverage of one set by another, or more formally ($I(U,V)$), is used as a metric. Thus, if one set of keys is completely included in the other for the collection pair $i$ and $j$, then the metric value will be equal to one, and $A_{ij} = 1$. If this value exceeds a pre-defined threshold, then the collections $i$ and $j$ are linked into a single entity. To solve the problem, we will consider the membership of one entity to be transitive, and since the matrix A is symmetric, it can be considered as an adjacency matrix for an undirected graph of collections with $n$ vertices. By reducing the matrix values to zero that do not exceed the specified metric threshold, the problem of combining collections is essentially reduced to searching of connected components of the graph.

All connected components can be found using the breadth-first search traversal algorithm (BFS) for each vertex, excluding previously added ones. Thus, an entity is represented by a set of its own collections, as well as a name generated based on the names of these collections. This approach allows you to incrementally add new collections to the data model -- just calculate which collections the new primary keys intersect with, and then add new vertices to the corresponding connected components. It should be noted that in this case, some components can be connected via a new collection, but combining the graph components is a trivial task, which does not introduce additional costs to the algorithm.

\begin{figure}[h]
    \subfloat[]{\label{subfig:num_cc_threshold}\includegraphics[width=0.5\textwidth]{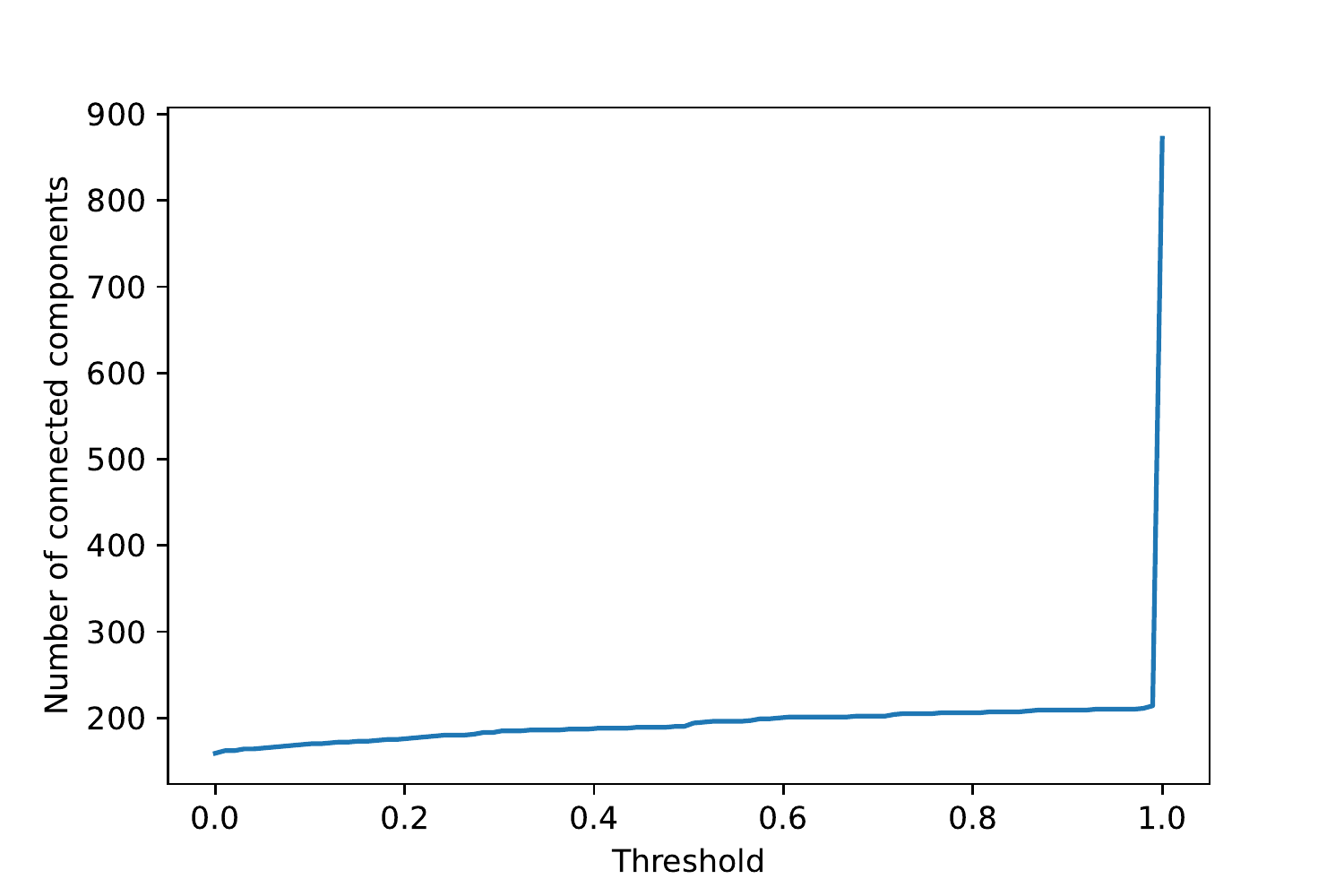}}
    \subfloat[]{\label{subfig:avg_cc_size_threshold}\includegraphics[width=0.5\textwidth]{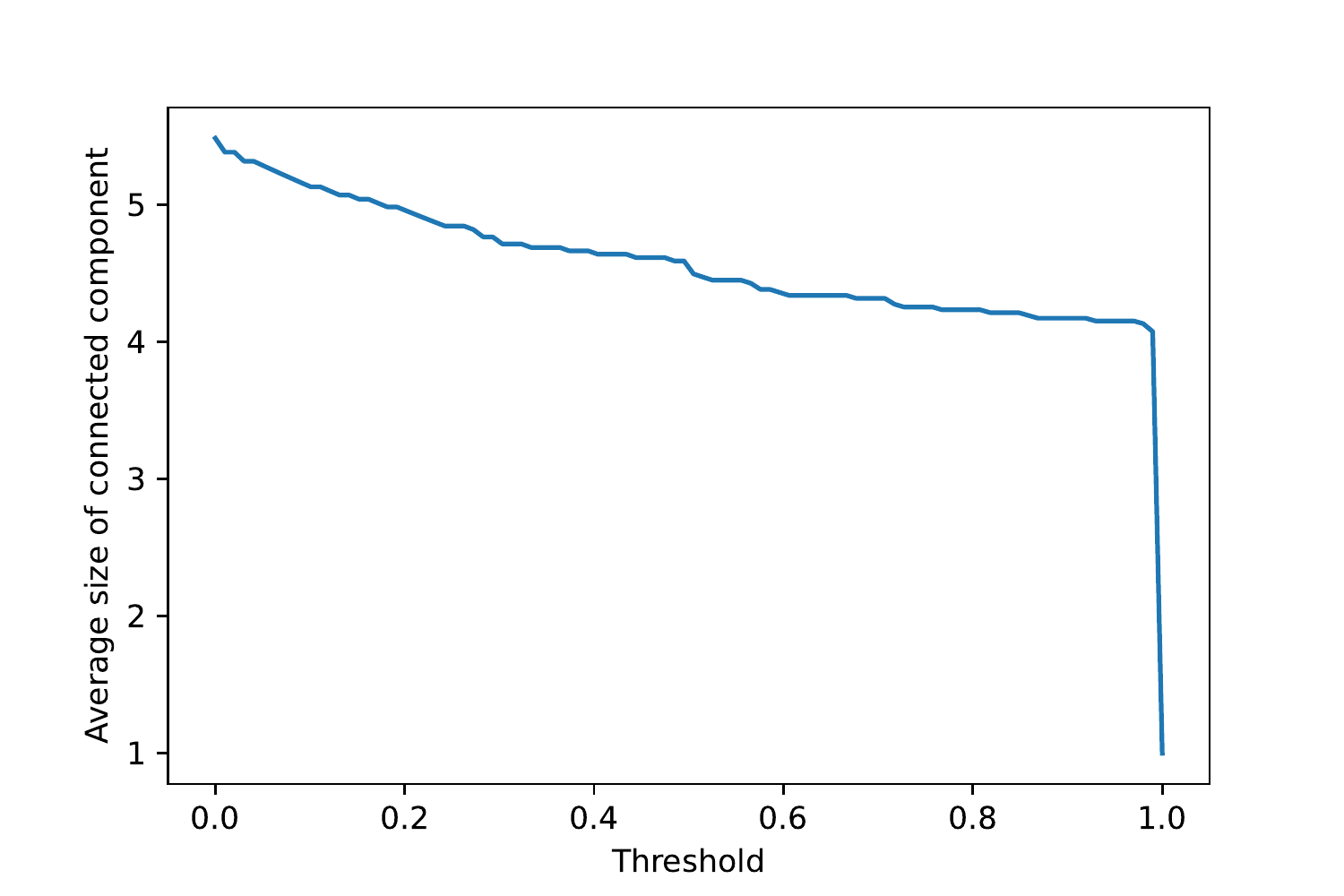}}
    \caption{Dependencies of the number of connected components (\ref{subfig:num_cc_threshold}) and the average size of the connected component (\ref{subfig:avg_cc_size_threshold}) on the metric threshold.}
    \label{fig:ccs_threshold}
\end{figure}

The dependencies of the number of entities (connected components) and the average size of entities on the threshold of the metric value are shown in Figure \ref{fig:ccs_threshold}.

\section{Framework architecture}
This section describes the software architecture of the methods and algorithms developed as a result of research. This architecture takes into account the sequence of execution of the developed components, and also describes the input and output data streams. Figure 6.1 shows a data flow diagram. The main input data for the operation of the developed software is the configuration file, as well as the external DBMS of the organization whose data needs to be processed. In the following sections, each component will be considered separately.

\begin{figure}
    \centering
    \includegraphics[width=\linewidth]{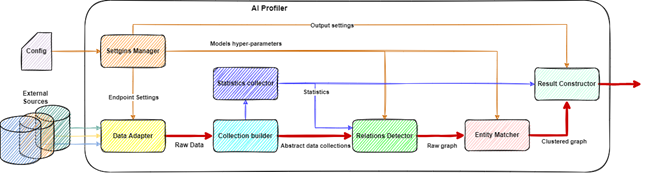}
    \caption{Data flow diagram. Red arrows indicate the main data stream, blue indicates the statistical data stream, and orange indicates parameters passed to software components.} \label{fig:schema}
\end{figure}

This component is the main input point of the developed software and is responsible for the mode of operation of other components. Configurations can be divided into three parts: 
\begin{itemize}
 \item   authorization parameters for external sources: contains descriptions of the protocols used, URIs, and information required to confirm access rights to the target DBMS.
 \item   hyper-parameters of the developed methods and algorithms.
\item parameters for the output data content and format.
 
\end{itemize}
The configuration file is described in YAML format with a top-level division into three logical parts.

\textbf{Data Source Adapter}
This component provides work with external DBMSs of the organization, provides other components with a single abstract interface for reading and processing data without binding to the specifics of the external DBMS. It accepts information about the protocol used and the data required for authorization as source data.

\textbf{Abstract Collection Linker}
	
This component provides uploading and primary processing of data from the organization's external DBMS using an abstract interface provided by the data source adapter. Primary data processing is performed using the method of filtering and preprocessing partially structured data (item 3). As a result, abstract collections that contain meta-information about the nature and structure of the source data are formed. The graph will be formed based on these collections. Additionally, this component collects global statistical characteristics of the organization's source data, which are necessary for building results or working of algorithms at subsequent stages of data processing, and transmits them to the statistics collection component.

\textbf{Statistics Collector}

This component provides collection and storage of global statistics necessary for the correct operation of the developed methods and algorithms, as well as for enriching the results with descriptive statistics and building analytics.

\textbf{Relationship Constructor}

This component is responsible for finding the relationship between abstract collections. Hyper-parameters of abstract collections and distribution of occurrence of unique identifiers (provided by statistics collector) are used as the input. The basis of this component is an algorithm of searching links between sources based on the data with any structure (**). The result of the work of this component is a graph whose nodes are abstract collections, and edges are facts of relations\textbackslash intersections between the collections. 

\textbf{Entity Constructor}

This component is responsible for combining abstract collections in an entity****** . hyperparameters and a graph of abstract collections are passed as input data.This component is based on the algorithm for combining collections in an entity (item 5). The result of this component is a graph where nodes are entities (combined collections), and edges are relations between them.

\textbf{Result Handler}

This component generates results in the form of an output file, in accordance with the configuration passed to the input. This component provides the ability to upload a mapping of field ownership in the source data structure to entities, as well as the history of changes in attributes of identified entities. This component is based on the "factory" design pattern, which allows you to safely change the format of the output file and the structure of the received data.

\section{Experimental setup}

The source data was provided by the customer as authorized access to the MongoDB document-oriented DBMS. Instead of tables, it stores collections – sets of documents, in turn, a document, unlike a record, can represent a complex nested structure of various data sets. A distinctive feature of this DBMS is the absence of any relationships between documents in collections, as well as its ability to store documents with different structures within the same collection, which complicates offline search for relationships.

Data is a set of collections, each of which contains a set of records from some source. A collection consists of a set of documents, each of which contains a record from a specific source, with its meta-information, such as the recording time, source, etc., while the main data of the record is located in the data field. There is also a collection that stores information about all collections, from which you can understand, what collections exist and how to access them.

In total, the data contains 895 collections, among which 894 contain at least one document. The distribution of the number of documents in collections is shown in Figure 2.1.1. In this distribution Q, Q1=7, Q2=40Q3, Q3=346, where QX is the corresponding quartile. In total, all collections contain 13584675 documents. In this case, the data field is of primary interest and is non-empty in 12928927 cases (95.17\%).
 

Each document in the collection contains information about a record of some source and its contents. The recording information is the recording time, source version, and other meta information, while the content is contained in the data field. The data field, in turn, is a json document that also contains fields that can be nested objects. The distribution of the degree of nesting of objects is shown in Figure . In turn, each object is either an array, or, if the object is a terminal node, a string, number, or other basic type. The distribution of node object types is shown in Figure 3.
 
 

In the course of exploratory data analysis, the distributions of the number of documents in collections, the degree of nesting of fields, and basic data types were considered. Data does not have a uniform structure, and the structure may vary from collection to collection. Moreover, the data structure may differ within the same collection, since the DBMS allows you to write objects of any type. Working with such data is also complicated by the fact that objects can be nested, as well as by the fact that about 29\% of the fields are not filled in. Also, collections are heterogeneous in terms of the number of documents in them, which potentially complicates their linking.

\section{Results}
When analyzing the source data, it was revealed that the data warehouse consists of 11 subsystems that contain 872 non-empty collections. After grouping collections with the same models and accounting for versioning, 430 collections were obtained. As a result of performing the procedures described in paragraphs 3-6, two graphs were constructed: 1) collection graph; 2) clustered graph (entity graph). Their characteristics are shown in Table 1.

\begin{figure}[h]
	\centering
	\subfloat[\label{sub:hours}]{\includegraphics[width=0.45\textwidth, height=5cm]{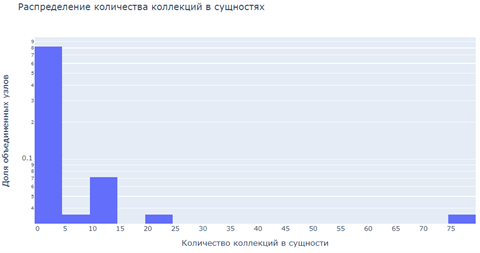}}\qquad
	\subfloat[\label{sub:weekdays}]{\includegraphics[width=0.45\textwidth, height=5cm]{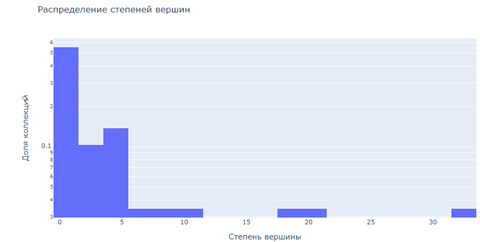}}\qquad
	\caption{Distribution of the number of collections in entities (left) and distribution of number links between entities (right)}
	\label{fig:ncoll}
\end{figure}

We can draw the following conclusions: both graphs are sparse, which correctly reflects the fact that the collection, as a rule, is connected only with a small number of other collections; a high modularity coefficient indicates that clusters representing different areas of activity can be distinguished.  

However, you can see that most collections are entities, and only a few entities are the product of combining many collections. This fact is illustrated by the distribution of the number of collections in entities (Figure 7.1).

It can be concluded that the resulting graphs link all 11 information systems of the customer. This is illustrated in Figure 7.2, which shows the number of edges between different subsystems. You can see that although the connectivity within a subsystem is higher, there are connections with other subsystems for each of them. Clustering the graph (on the right) reduces the noise of connections, making the structure more explicit.
 	 

Thus, as a result of the work, the task of constructing a domain-specific knowledge graph based on customer data was completed, and it covered all 11 subsystems of data storage, while remaining very sparse – there are only 2.5 connections on average.

\section{Discussion and conclusion}

We developed a method for filtering and preprocessing of partially structured data, and aggregating it into data structures, taking into account the possibility of changing the set of sources and their fields over time.
We developed an algorithm for finding links between sources based on data with an arbitrary structure in automatic mode and a method for integrating structured and partially structured data into graphs, where nodes are tables or collections of source databases, and edges reflect the fact of relationships/intersections between data entities of tables/collections. 
Based on the proposed methodology and algorithms, a software library for automatic profiling and building an organization's knowledge graph in Python 3.9 was implemented.
The developed library was tested on customer data, which showed the applicability of the developed approach for building a knowledge graph based on partially structured data from various sources and high efficiency in terms of completeness of data storage subsystems coverage (11 out of 11) and filtering false connections (in the final graph, there are only 2.5 connections of collections with neighbors on average).
Thus, in the course of the work, the goal was achieved – the development of methods and algorithms for automatically compiling and enriching a domain-specific knowledge graph based on intelligent profiling of partially structured industrial data.

\section*{\uppercase{Acknowledgements}}
This research is financially supported by the Ministry of Science and Higher Education, Agreement FSER-2021-0012

\bibliographystyle{unsrtnat}
\bibliography{mybib}  

\begin{thebibliography}{25}
\providecommand{\natexlab}[1]{#1}
\providecommand{\url}[1]{\texttt{#1}}
\expandafter\ifx\csname urlstyle\endcsname\relax
  \providecommand{\doi}[1]{doi: #1}\else
  \providecommand{\doi}{doi: \begingroup \urlstyle{rm}\Url}\fi

\bibitem[Hogan et~al.(2021)Hogan, Blomqvist, Cochez, D'Amato, Melo, Gutierrez,
  Kirrane, Gayo, Navigli, Neumaier, Ngomo, Polleres, Rashid, Rula,
  Schmelzeisen, Sequeda, Staab, and Zimmermann]{Hogan2021}
Aidan Hogan, Eva Blomqvist, Michael Cochez, Claudia D'Amato, Gerard~De Melo,
  Claudio Gutierrez, Sabrina Kirrane, Jos{\'{e}} Emilio~Labra Gayo, Roberto
  Navigli, Sebastian Neumaier, Axel Cyrille~Ngonga Ngomo, Axel Polleres,
  Sabbir~M. Rashid, Anisa Rula, Lukas Schmelzeisen, Juan Sequeda, Steffen
  Staab, and Antoine Zimmermann.
\newblock {Knowledge graphs}.
\newblock \emph{ACM Computing Surveys}, 54\penalty0 (4), 2021.
\newblock ISSN 15577341.
\newblock \doi{10.1145/3447772}.

\bibitem[Cur{\'{e}} et~al.(2013)Cur{\'{e}}, Lamolle, and Duc]{Cure2013}
Olivier Cur{\'{e}}, Myriam Lamolle, and Chan~Le Duc.
\newblock {Ontology Based Data Integration Over Document and Column Family
  Oriented NOSQL}.
\newblock 2013.

\bibitem[Abbes et~al.(2015)Abbes, Boukettaya, and Gargouri]{Abbes2015}
Hanen Abbes, Soumaya Boukettaya, and Faiez Gargouri.
\newblock {Learning ontology from Big Data through MongoDB database}.
\newblock In \emph{2015 IEEE/ACS 12th International Conference of Computer
  Systems and Applications (AICCSA)}, pages 1--7, 2015.
\newblock \doi{10.1109/AICCSA.2015.7507166}.

\bibitem[Abbes and Gargouri(2017)]{Abbes2017}
Hanen Abbes and Faiez Gargouri.
\newblock {M2Onto: An Approach and a Tool to Learn OWL Ontology from MongoDB
  Database}.
\newblock In Ana~Maria Madureira, Ajith Abraham, Dorabela Gamboa, and Paulo
  Novais, editors, \emph{Intelligent Systems Design and Applications}, pages
  612--621, Cham, 2017. Springer International Publishing.
\newblock ISBN 978-3-319-53480-0.

\bibitem[Sbai et~al.(2019)Sbai, Louhdi, Behja, and Chakhmoune]{Sbai2019}
Sara Sbai, Mohammed Reda~Chbihi Louhdi, Hicham Behja, and Rabab Chakhmoune.
\newblock {JsonToOnto: Building Owl2 Ontologies from Json Documents}.
\newblock \emph{International Journal of Advanced Computer Science and
  Applications}, 10\penalty0 (10), 2019.
\newblock \doi{10.14569/IJACSA.2019.0101030}.

\bibitem[Lample et~al.(2016)Lample, Ballesteros, Subramanian, Kawakami, and
  Dyer]{Lample2016}
Guillaume Lample, Miguel Ballesteros, Sandeep Subramanian, Kazuya Kawakami, and
  Chris Dyer.
\newblock {Neural Architectures for Named Entity Recognition}.
\newblock In \emph{Proceedings of the 2016 Conference of the North
  {\{}A{\}}merican Chapter of the Association for Computational Linguistics:
  Human Language Technologies}, pages 260--270, San Diego, 2016. Association
  for Computational Linguistics.
\newblock \doi{10.18653/v1/N16-1030}.

\bibitem[Gupta and Manning(2014)]{Gupta2014}
Sonal Gupta and Christopher Manning.
\newblock {Improved Pattern Learning for Bootstrapped Entity Extraction}.
\newblock In \emph{Proceedings of the Eighteenth Conference on Computational
  Natural Language Learning}, pages 98--108, Ann Arbor, 2014. Association for
  Computational Linguistics.
\newblock \doi{10.3115/v1/W14-1611}.

\bibitem[Nakashole et~al.(2013)Nakashole, Tylenda, and Weikum]{Nakashole2013}
Ndapandula Nakashole, Tomasz Tylenda, and Gerhard Weikum.
\newblock {Fine-grained Semantic Typing of Emerging Entities}.
\newblock In \emph{Proceedings of the 51st Annual Meeting of the Association
  for Computational Linguistics (Volume 1: Long Papers)}, pages 1488--1497,
  Sofia, 2013. Association for Computational Linguistics.

\bibitem[Yogatama et~al.(2015)Yogatama, Gillick, and Lazic]{Yogatama2015}
Dani Yogatama, Daniel Gillick, and Nevena Lazic.
\newblock {Embedding Methods for Fine Grained Entity Type Classification}.
\newblock In \emph{Proceedings of the 53rd Annual Meeting of the Association
  for Computational Linguistics and the 7th International Joint Conference on
  Natural Language Processing (Volume 2: Short Papers)}, pages 291--296,
  Beijing, 2015. Association for Computational Linguistics.
\newblock \doi{10.3115/v1/P15-2048}.

\bibitem[Ling and Weld(2012)]{Ling2012}
Xiao Ling and Daniel Weld.
\newblock {Fine-Grained Entity Recognition}, 2012.

\bibitem[Ren et~al.(2015)Ren, El-Kishky, Wang, Tao, Voss, Ji, and Han]{Ren2015}
Xiang Ren, Ahmed El-Kishky, Chi Wang, Fangbo Tao, Clare~R Voss, Heng Ji, and
  Jiawei Han.
\newblock {ClusType: Effective Entity Recognition and Typing by Relation
  Phrase-Based Clustering}.
\newblock In \emph{Proceeding of 2015 ACM SIGKDD International Conference on
  Knowledge Discovery and Data Mining}. ACM – Association for Computing
  Machinery, 2015.

\bibitem[Chiticariu et~al.(2018)Chiticariu, Danilevsky, Li, Reiss, and
  Zhu]{Chiticariu2018}
Laura Chiticariu, Marina Danilevsky, Yunyao Li, Frederick Reiss, and Huaiyu
  Zhu.
\newblock {SystemT: Declarative Text Understanding for Enterprise}.
\newblock In \emph{Proceedings of the 2018 Conference of the North
  {\{}A{\}}merican Chapter of the Association for Computational Linguistics:
  Human Language Technologies, Volume 3 (Industry Papers)}, pages 76--83, New
  Orleans - Louisiana, 2018. Association for Computational Linguistics.
\newblock \doi{10.18653/v1/N18-3010}.

\bibitem[Kluegl et~al.(2009)Kluegl, Atzmueller, and Puppe]{Kluegl2009}
Peter Kluegl, Martin Atzmueller, and Frank Puppe.
\newblock {TextMarker: A Tool for Rule-Based Information Extraction}.
\newblock In \emph{Proc. Unstructured Information Management Architecture
  (UIMA), 2nd UIMA@GSCL Workshop, 2009 Conference of the GSCL (Gesellschaft
  f{\"{u}}r Sprachtechnologie und Computerlinguistik)}, 2009.

\bibitem[Chiticariu et~al.(2013)Chiticariu, Li, and Reiss]{Chiticariu2013}
Laura Chiticariu, Yunyao Li, and Frederick~R Reiss.
\newblock {Rule-Based Information Extraction is Dead! Long Live Rule-Based
  Information Extraction Systems!}
\newblock In \emph{Proceedings of the 2013 Conference on Empirical Methods in
  Natural Language Processing}, pages 827--832, Seattle, 2013. Association for
  Computational Linguistics.

\bibitem[Moro et~al.(2014)Moro, Raganato, and Navigli]{Moro2014}
Andrea Moro, Alessandro Raganato, and Roberto Navigli.
\newblock {Entity Linking meets Word Sense Disambiguation: a Unified Approach}.
\newblock \emph{Transactions of the Association for Computational Linguistics},
  2:\penalty0 231--244, 2014.
\newblock \doi{10.1162/tacl_a_00179}.

\bibitem[Wu et~al.(2018)Wu, He, and Hu]{Wu2018}
Gongqing Wu, Ying He, and Xuegang Hu.
\newblock {Entity Linking: An Issue to Extract Corresponding Entity With
  Knowledge Base}.
\newblock \emph{IEEE Access}, 6:\penalty0 6220--6231, 2018.
\newblock \doi{10.1109/ACCESS.2017.2787787}.

\bibitem[Roller et~al.(2018)Roller, Kiela, and Nickel]{Roller2018}
Stephen Roller, Douwe Kiela, and Maximilian Nickel.
\newblock {Hearst Patterns Revisited: Automatic Hypernym Detection from Large
  Text Corpora}.
\newblock In \emph{Proceedings of the 56th Annual Meeting of the Association
  for Computational Linguistics (Volume 2: Short Papers)}, pages 358--363,
  Melbourne, 2018. Association for Computational Linguistics.
\newblock \doi{10.18653/v1/P18-2057}.

\bibitem[Bunescu and Mooney(2005)]{Bunescu2005}
Razvan~C Bunescu and Raymond~J Mooney.
\newblock {Subsequence Kernels for Relation Extraction}.
\newblock In \emph{Proceedings of the 18th International Conference on Neural
  Information Processing Systems}, NIPS'05, pages 171--178, Cambridge, 2005.
  MIT Press.

\bibitem[Etzioni et~al.(2004)Etzioni, Cafarella, Downey, Kok, Popescu, Shaked,
  Soderland, Weld, and Yates]{Etzioni2004}
Oren Etzioni, Michael Cafarella, Doug Downey, Stanley Kok, Ana-Maria Popescu,
  Tal Shaked, Stephen Soderland, Daniel~S Weld, and Alexander Yates.
\newblock {Web-Scale Information Extraction in Knowitall: (Preliminary
  Results)}.
\newblock In \emph{Proceedings of the 13th International Conference on World
  Wide Web}, WWW '04, pages 100--110, New York, 2004. Association for Computing
  Machinery.
\newblock ISBN 158113844X.
\newblock \doi{10.1145/988672.988687}.

\bibitem[Smirnova and Cudr{\'{e}}-Mauroux(2018)]{Smirnova2018}
Alisa Smirnova and Philippe Cudr{\'{e}}-Mauroux.
\newblock {Relation Extraction Using Distant Supervision: A Survey}.
\newblock \emph{ACM Comput. Surv.}, 51\penalty0 (5), 2018.
\newblock ISSN 0360-0300.
\newblock \doi{10.1145/3241741}.

\bibitem[Nentwig et~al.(2017)Nentwig, Hartung, {Ngonga Ngomo}, and
  Rahm]{Nentwig2017}
Markus Nentwig, Michael Hartung, Axel-Cyrille {Ngonga Ngomo}, and Erhard Rahm.
\newblock {A survey of current Link Discovery frameworks}.
\newblock \emph{Semantic Web}, 8:\penalty0 419--436, 2017.
\newblock ISSN 2210-4968.
\newblock \doi{10.3233/SW-150210}.

\bibitem[Sun et~al.(2020)Sun, Zhang, Hu, Wang, Chen, Akrami, and Li]{Sun2020}
Zequn Sun, Qingheng Zhang, Wei Hu, Chengming Wang, Muhao Chen, Farahnaz Akrami,
  and Chengkai Li.
\newblock {A benchmarking study of embeddingbased entity alignment for
  knowledge graphs}.
\newblock \emph{Proceedings of the VLDB Endowment}, 13\penalty0 (11):\penalty0
  2326--2340, 2020.
\newblock ISSN 21508097.
\newblock \doi{10.14778/3407790.3407828}.

\bibitem[Zhao et~al.(2020)Zhao, Zeng, Tang, Wang, and Suchanek]{Zhao2020}
Xiang Zhao, Weixin Zeng, Jiuyang Tang, Wei Wang, and Fabian Suchanek.
\newblock {An Experimental Study of State-of-the-Art Entity Alignment
  Approaches}.
\newblock \emph{IEEE Transactions on Knowledge and Data Engineering}, 2020.
\newblock ISSN 15582191.
\newblock \doi{10.1109/TKDE.2020.3018741}.

\bibitem[Konda et~al.(2016)Konda, Das, {Suganthan G. C.}, Doan, Ardalan,
  Ballard, Li, Panahi, Zhang, Naughton, Prasad, Krishnan, Deep, and
  Raghavendra]{Konda2016}
Pradap Konda, Sanjib Das, Paul {Suganthan G. C.}, AnHai Doan, Adel Ardalan,
  Jeffrey~R Ballard, Han Li, Fatemah Panahi, Haojun Zhang, Jeff Naughton,
  Shishir Prasad, Ganesh Krishnan, Rohit Deep, and Vijay Raghavendra.
\newblock {Magellan: Toward Building Entity Matching Management Systems}.
\newblock \emph{Proc. VLDB Endow.}, 9\penalty0 (12):\penalty0 1197--1208, 2016.
\newblock ISSN 2150-8097.
\newblock \doi{10.14778/2994509.2994535}.

\bibitem[Mudgal et~al.(2018)Mudgal, Li, Rekatsinas, Doan, Park, Krishnan, Deep,
  Arcaute, and Raghavendra]{Mudgal2018}
Sidharth Mudgal, Han Li, Theodoros Rekatsinas, AnHai Doan, Youngchoon Park,
  Ganesh Krishnan, Rohit Deep, Esteban Arcaute, and Vijay Raghavendra.
\newblock {Deep Learning for Entity Matching: A Design Space Exploration}.
\newblock In \emph{Proceedings of the 2018 International Conference on
  Management of Data}, SIGMOD '18, pages 19--34, New York, 2018. Association
  for Computing Machinery.
\newblock ISBN 9781450347037.
\newblock \doi{10.1145/3183713.3196926}.

\end{thebibliography}

\end{document}